\documentclass[%
 aip,
 amsmath,amssymb,
 reprint,%
]{revtex4-1}

\usepackage{graphicx}
\usepackage{dcolumn}
\usepackage{bm}

\usepackage[utf8]{inputenc}
\usepackage[T1]{fontenc}
\usepackage{mathptmx}
\usepackage{etoolbox}
\usepackage{hyperref}

\makeatletter
\def\@email#1#2{%
 \endgroup
 \patchcmd{\titleblock@produce}
  {\frontmatter@RRAPformat}
  {\frontmatter@RRAPformat{\produce@RRAP{*#1\href{mailto:#2}{#2}}}\frontmatter@RRAPformat}
  {}{}
}%
\makeatother
\begin{document}

\preprint{AIP/123-QED}

\title[Oral Exams]{Using Oral Exams in Physics and Astronomy Courses}

\author{Brian DiGiorgio Zanger}
\altaffiliation[Previously at ]{Department of Physics \& Engineering Physics, Juniata College, 1700 Moore St., Huntingdon, PA 16652, USA}
\affiliation{Physics and Astronomy Department, Swarthmore College, 500 College Avenue, Swarthmore, PA 19081, USA}
\email{bzanger1@swarthmore.edu}

\date{\today}

\begin{abstract}
    Oral exams were common historically across academia, though their popularity has recently fallen. Many argue against them as an assessment technique because they are vulnerable to bias and subjectivity, difficult to administer, and impractical in large college classes. I present a method for administering oral examinations in upper-level courses that mitigates some of the disadvantages. This method creates a rigid question structure meant to assess student mastery of material, have a well-defined grading structure to standardize evaluation, and be administered a constrained time limit to reduce workload in larger seminars. I emphasize holistic verbal communication and evaluation that is meant to mirror the talks and interviews that are common throughout a scientist's career.
\end{abstract}

\maketitle

\section{Introduction}
Oral examinations are some of the oldest forms of assessment in academia. Beginning hundreds of years ago in the era of one-on-one tutoring, \textit{viva voce} exams provided an individually-tailored method for a mentor to probe the limits of their student's knowledge \citep{pearce09}. However, beyond  doctoral qualification exams and dissertation defenses, the oral exam has largely disappeared from modern physics and astronomy in the United States, leaving most professors to rely on written exams for assessment \citep{ehrlich07}. In recent decades, oral exams have been criticized for being time-consuming to administer \citep{kang22}, uneven between students \citep{platt61}, and susceptible to bias from the examiner \citep{burke14}.

Despite these misgivings, oral exams continue to be practiced sporadically across other fields. \citet{luckie13} describes an hour-long verbal final exam to students in an introductory biology course, using the extended timing to ask students to explain long and complex biological mechanisms. \citet{kamber21} evaluates students based on the confidence of their answer in an undergraduate biochemistry class, and finds that oral exams are effective in remote learning environments. \citet{kang22} deputizes graduate student TAs to perform hundreds of short oral exams for an undergraduate humanities course. \citet{burke14} finds that oral exams allow students to practice verbal communication skills necessary for a career in business. \citet{mariano24} notes that oral exams resist cheating attempts with AI chatbots like ChatGPT, Gemini, or Claude. These chatbots have higher usage among natural sciences students than in many other fields, and students have demonstrated a willingness to disproportionately offload higher-order cognitive tasks to AI \citep{claude25}, hampering learning and leading many institutions to search for alternative assessment methods \citep{shirky25}.

Though oral examination reportedly remains more common in other countries like Italy \citep{ehrlich07} and Denmark \citep{hurford20}, literature on  physics and astronomy oral exams within the United States is more limited and dated. \citet{styer97} used a 30 minute oral exam as one part of a larger lesson plan within a statistical mechanics course, balancing the analytic skills of the problem sets with the conceptual environment of the oral exam. \citet{ehrlich07} and \citet{boedecker78} allowed students to earn extra points after a written exam by verbally answering further questions about problems that they missed.

In this paper, I detail my method for administering oral exams, which has received generally positive student reactions, and I enumerate its pedagogical benefits. I also address a number of drawbacks and biases in oral examination and explain how my method attempts to mitigate them.

\section{Oral Exam Method} \label{method}
Below is a summary of my oral exam procedures. I have performed these procedures in a range of upper-level physics and astronomy courses at two separate small liberal arts colleges, with class sizes ranging from 7-15 students. In principle, these methods could be applied in courses with different sizes or in different institutional environments, though there are limitations on practicality that become important with larger class sizes. 

Students are provided a sheet explaining these policies and procedures ahead of time, which can be found in the online supplementary materials.

\subsection{Exam structure}
The instructor prepares a set of question categories that span the material covered in the assessment. In each question category, three different questions are prepared: an ``explanation question," an ``application question," and an ``extension question." These question types are chosen to assess mastery, as defined in \citet{hlw}. In their formalism, students first acquire component skills (assessed by the explanation question), then they integrate those skills into a broader problem (application question), before finally transferring them to a new context (extension question). If a student successfully answers all of these questions, then they have demonstrated mastery of the subject. These criteria also mirror Bloom's taxonomy of education goals, with later questions reaching higher taxonomic levels to assess learning \citep{hurford20}.

A more detailed application to physics problems and example questions are below. Further examples can be found in Appendix \ref{examples}.
\begin{itemize}
    \item \textbf{Explanation question:} The first question asks the student to explain some relevant section of the material in a short lecture and often drawing an accompanying diagram. This question is meant to assess the student's baseline knowledge of the material. For example, in a modern physics course, an explanation question in a category about the Rutherford gold foil experiment may be ``Draw a diagram of the Rutherford experiment apparatus and explain why the different parts were chosen and arranged this way. What was the unexpected observation Rutherford made?" A successful student would reproduce the diagram and talk about the observed back-scatter.
    \item \textbf{Application question:} This question requires the student to apply the surface-level knowledge of the problem they just articulated to an actual problem. This is meant to assess the depth of their knowledge and their ability to solve a problem easily, and it may require interpreting an equation or principle from the explanation question that goes beyond basic information that they could easily memorize. For the Rutherford category, this may be ``The plum pudding model has dense concentrations of charge within the atom (the electrons). Why didn’t Rutherford expect that these electrons would deflect the particle beam very much?" A successful answer would use either conceptual arguments or the Rutherford scattering equation to explain the connection between the electron's mass and charge and its ability to scatter particles at large angles.
    \item \textbf{Extension question:} This question invites the student to extend their knowledge beyond the material that was explicitly covered and to make new connections. Often, these questions rely on some example from another course, an application of the material in the real world, or a hypothetical modification to the system that would yield different results. The purpose of this question is to assess the student's mastery by seeing how they can connect disparate subjects and incorporate new information into their understanding of the course material. An example for the Rutherford experiment would be ``Superheavy elements (the artificially constructed ones at the bottom of the periodic table) are often made in particle accelerators, where heavy nuclei like calcium are fired at a metal target at very high speeds. This setup looks qualitatively similar to the Rutherford gold foil experiment. Why does it yield a different result, and why is it important to do these experiments in particle accelerators that can get the nuclei going extremely fast?" A successful answer would connect that at very high energies, the particles can approach the nucleus close enough to fuse.
\end{itemize}

I typically ask conceptual rather than mathematical questions since students often take very different amounts of time to solve equations and may experience more anxiety when doing mathematics in a live setting. I instead rely on problem sets for assessment of a student's mathematical skill since they can be completed at a student's own pace. Other instructors may choose to have an oral exam as a supplement to a written exam \citep{ehrlich07} or to include mathematical skills in oral exams to replace or supplement problem sets, but I choose a complementary role to diversify the assessment methods in the course, as suggested by the principles of Universal Design for Learning \citep{udl}.

In order to minimize the sharing of question information between students, the instructor should prepare more question categories than are necessary for a single exam, and question categories should be chosen randomly for each student. A single student should only see a fraction of the total material prepared, and it should be rare for any two students to experience identical exams. For example, when I administer an oral midterm exam, I prepare six or eight categories and only have students complete two of them. Instructors can also control a student's access to outside materials much more closely in a one-on-one environment, reducing cheating and AI use \citep{mariano24}.

\subsection{Procedure}
Students are invited to schedule a timeslot for their exam over a period of three or four days. A more detailed description of logistics is described in informational sheet that can be found in the online supplementary materials for this article and which is distributed to students beforehand.

During their timeslot, the student comes to a meeting in a suitable private space. The student is presented with a list of categories on a sheet of paper. I do not provide these categories ahead of time to encourage students to study more broadly, but other instructors may choose differently. For a midterm exam, I require students to complete two full question categories (six total questions) over the course of 20 minutes (10 minutes per category). Oral exams can cause a great deal of anxiety at first in some students because they are not accustomed to being assessed in this manner \citep{kamber21}, so I attempt to return some agency to the students by allowing them to choose one of their question categories themselves. For the other category (or categories), I require them to roll a die to randomly pick between categories to minimize cheating risk, using a six- or eight-sided die depending on the number of categories. \footnote{Students with exceptional anxiety can be granted additional \textit{ad hoc} accommodations depending on institutional rules. \citet{hurford20} also suggests integrating a peer-administered practice exam into the normal class period to acclimate students to the format and also see the perspective of the examiner.}

I then provide a half sheet of paper that has all of the questions for the category, reading each question aloud as it becomes relevant so students can access the information in the manner they prefer. Students are permitted to answer verbally or on a whiteboard, depending on their preference or on the nature of the question. When a student has completed their questions, I quickly debrief the student's performance and feelings on the exam before releasing them.

I find that the time pressure of 10 minutes per category is useful for forcing students to answer relatively quickly while still allowing over three minutes per question on average. I also try to build in a few minutes of extra time to allow students to go over time slightly and to facilitate transitions between students in back-to-back meetings. The maximum class size in which I have used this system is 15 students, which corresponded to a total of five hours of student meetings. 

For a final exam, I retain the same procedures but with three categories instead of two. I reuse categories and questions from my midterm, combining all of the questions into one master list of categories that requires students to roll a 20-sided die. Before the exam, I compile a list of which categories each student has already completed to prevent any students from gaining an unfair advantage by repeating material.

\subsection{Grading and Bias} \label{grading}
One of the most commonly cited disadvantages of oral exams is their tendency to be subjective and variable \citep{platt61}. Traditionally, oral exams are graded based on an instructor's subjective assessment of a student's understanding \citep[\textit{e.g.}][]{ehrlich07, pearce09, boedecker78}, creating opportunities for bias to enter into student grades \citep{mariano24}. Instructors may experience stereotype bias (different student expectations based on perceived group membership) or halo bias (different student expectations based on past performance or personal relationships), both of which are more easily mitigated in an anonymized written exam than in an in-person oral exam \citep{gerlt23}. Therefore, it is necessary to define a standardized scoring system before beginning the assessment that is generalizable to each of the questions. 

I allocate 10 points to each of the questions, making each category worth a total of 30 points. For each question, students start with a perfect 10/10 points, but each time I intervene, they lose a point. These interventions usually take the form of subtle hints to point them in the correct problem-solving direction, corrections of any incorrect statements they have made, or prompting on some information they omitted. Because student mistakes are idiosyncratic, hints are often difficult to plan ahead of time, so the instructor must take care not to provide some students with more information than others. The interventions continue until a student produces a correct answer (or, rarely, until no more hints can be provided without outrightly stating the answer), so more major errors are naturally punished more than smaller mistakes due to those errors likely coming from larger underlying misunderstandings. This direct and prompt feedback has been shown to increase the efficiency with which student develop mastery in a subject \citep{hlw}.

Using the example of Rutherford's experiment from above, if a student does not address why gold was chosen as a target material, I will prompt them by asking that question separately, or if they incorrectly say that an electron deflects the alpha particle beam less than the nucleus because of its diameter rather than its charge, then I will call out this mistake specifically and ask them to reconsider. If these prompts do not elicit a correct response, then more specific Socratic-style questions can be asked to lead the student to the correct answer (``What properties would a gold nucleus have based on its place in the periodic table?", ``What other ways do electrons and gold nuclei differ?"). Each additional question of this style will cost the student an additional point.

I try to be as permissive as is practical in letting students find their own way and catch their own mistakes, only intervening as a last resort or when time is short. I do not penalize a student who is able to correct themselves eventually without intervention (unlike \citet{kamber21}) or if a student asks a clarifying question. I also do not penalize for any incorrect statements that are unrelated to the question since that is not the information being assessed and such penalties may bias against students who give more verbose responses, though the instructor may wish to correct these mistakes anyways.

Care must be taken to make each category approximately the same difficulty since not all students will encounter the same questions in their exam. If one category is judged to be easier or harder than others when testing on real students (for example, students consistently struggle with the answer even after prompting), then the question can be adjusted between sessions and/or the scores can be updated in retrospect, but this may be difficult to determine objectively. 

Questions must be formulated to have a clearly defined answer to reduce subjective areas of bias \citep{gerlt23}. Traditional oral exams reward students for in-depth answers beyond the surface level, but evaluation based on open-ended questions and the depth of the answer invites more opportunities for bias than a single correct answer regardless of depth \citep{hurford20}.  In addition, allowing students to choose one of their question categories relies on the student's assessment of their own knowledge and whether they can accurately anticipate what will be asked in a given category, a bias that can be mitigated by removing student choice.

In general, a student with a great understanding of the topic will need only one or zero interventions and a poorly-prepared student will require five or more, with average students falling in between. This usually results in a scoring distribution with a mean in the high 80\% range, with perfect scores being exceptionally rare and a long left tail for unprepared students, though scoring distributions will vary with question design and student population. I total up the individual question scores to communicate the final grade at the end of the exam. Because grading is done in real time, the overall grading burden for the exam is significantly lower than a typical written exam.

\subsection{Student Response}
I have now employed this method in three different courses so far, and below I have summarized the major themes in student feedback across those courses. Quotations in italics are pulled directly from informal student feedback when asked generally about their thoughts on the oral exams, how they went, and how they could be changed.

\begin{itemize}
\item \textit{``The oral midterm is a very good idea"}: Broadly, almost all students came out feeling positively about the exams even after their first experience. They recognized the utility of the method (\textit{``The oral exam is very helpful in developing the skills of explaining my knowledge to others, which is important as a physicist"}) and felt that it added to their experience in the course (\textit{``I feel like I had to study much harder (in a good way) in order to do well"}).

\item \textit{``It wasn't as terrible as I feared"}: Many expressed initial anxiety about the unfamiliar format, though most student concerns are quickly assuaged by going through the process once. Subsequent student feedback was much more positive (\textit{``I knew what to expect so they were fine"}).

\item \textit{``Less time pressure would be nice"}: Some students felt rushed, leading to increased stress. In particular, non-native English speakers reported feeling additional pressure (\textit{``I am not native US so when the time I get the question is quite hard for me [sic]"}), potentially warranting broader accommodations. Students also expressed frustration and anxiety about the randomness of question selection and not having access to categories ahead of time, so other instructors may choose to modify my method to address these concerns.

\item \textit{``Oral midterms feels like a conversation"}: Some students report feeling more relaxed in the oral exam compared to written exams, but some find difficulty in expressing their thoughts in a new medium at first (\textit{``I struggle to vocalize my thought process"}).

\item \textit{``I got the best score I have ever gotten on any college-level exam on this midterm, and I think I entirely owe it to the fact that I didn't have to sit down in a silent room for 2 hours."}: Once students are comfortable with the format, many find that it accommodates their learning differences more easily, though some students with anxiety remained uncomfortable (\textit{``I have anxiety so it was a little bit terrifying and at times I felt like I couldn't take the time to really think"}). As always, learning differences warrant specific consultation with the student to best accommodate them.
\end{itemize}

\section{Benefits and drawbacks of oral exams} \label{procon}
\subsection{Benefits}

\textbf{Increased communication confidence:} Scientists are almost never assessed in a timed, closed note, written environment in their professional lives. Rather, they are often evaluated by their ability to explain a complex idea orally to an audience, like in a conference talk, job interview, or presentation to a client \citep{ehrlich07}. Thus, oral exams follow the principles of backward design by assessing students in the manner they will be expected to demonstrate knowledge in the future. Past studies on oral exams have found improvements in professional oral communication ability \citep{burke14}, and I have personally observed many students who become much more adept at dealing with the pressure of oral presentation with practice throughout the semester.

\textbf{Greater question freedom:} In a traditional written exam, students who do not know how to answer a problem may have no options besides guessing or moving on. Complex problems on exams may be more effective in evaluating finer points of a subject but may yield lower student grades, while simpler problems may yield better grades but not separate students by expertise as effectively. However, a live examiner in an oral exam can ask a complex question and provide gentle guidance or hints to struggling students, allowing them to make new connections within the material, and frequent, prompt feedback accelerates a student's development of expertise \citep{hlw}.

\textbf{Flexibility for accommodations:} Because oral exams are administered individually, each one can be administered in the form that makes the most sense for the student. Individual exams are distraction-free, and timing and venue can easily be varied based on student need. In addition, oral exams lend themselves well to being administered remotely through a video conferencing application \citep{kamber21}.

\textbf{Reduced cheating risk:} Overall, cheating is much more difficult in an oral exam environment due to closer instructor oversight. However, there is the new possibility of students sharing questions between exams, which can be ameliorated by a larger question pool, by switching out specific questions if they have been chosen too many times, or limiting student interaction between sessions.

\textbf{Holistic assessment:} In my experience, watching a student solve a problem in real time gives a much more detailed view of their work habits than grading a written assessment. Students who struggle with complex mathematics may show exceptional capability with forming new conceptual connections, and students who often get stuck in isolation may only require one nudge to complete a problem. No single assessment can be perfect, and assessing in a different manner gives a new data point on student competence and provides a different window into a student's understanding, a practice encouraged by Universal Design for Learning \citep{udl}.

\subsection{Drawbacks}

\textbf{Time commitment:} Oral exams require significant time to administer, which may be a burden on an instructor's schedule. As mentioned in Section \ref{method}, I have only ever applied this method in liberal arts college environments for classes of 7-15 students, and scaling up to larger class sizes may be prohibitive. Time constraints can be lifted somewhat by deputizing student TAs to serve as examiners as well \citep{kang22} or by administering exams in groups \citep{guest00}.

\textbf{Limited exam length:} The practicalities of administering individual oral exams places time pressure on how long a given exam can be. Longer assessments allow for more individual measurements of student competence \citep{ellis21}, so a shorter exam with limited scope may lead to more student grades that are erroneously high or low due to the incomplete sampling of the course material and the student's knowledge.

\textbf{Verbal presentation difficulties:} Some students experience large amounts of anxiety when being evaluated in an isolated environment, artificially hurting their scores due to stress. Though some of this can be accommodated by changes in exam environment, it may lead to disparate exam outcomes. In addition, students with language processing disorders or non-native speakers may have difficulties in parsing the nuanced phrasing of a question or hint in real time, leading to potential disadvantages compared to written tests \citep{hurford20}.

\textbf{Difficult to administer:} In addition to student discomfort, instructors are also more accustomed to creating and administering written examinations, so running oral examinations may be difficult for an instructor to learn and perfect, further increasing workload and mental effort. Creating a large question bank of equal difficulty questions may also be taxing.

\section{Conclusion}
Given their benefits as an assessment tool, oral exams merit reconsideration in a modern learning environment. They may require more effort and time investment on the part of the instructor, but they emphasize skills in verbal communication, foster connections across disparate areas of the material, and present a more holistic view of a student's learning. My method is designed around questions that directly assess student expertise in the subject, while mitigating traditional disadvantages of subjectivity, variability, and bias by creating a standard question format and grading scheme. Oral exams also reduce AI cheating risk, increase accessibility, and adapt well to remote classes in a time when those needs are more pressing. Though oral exams are likely impractical for large classes, instructors of smaller seminar courses may find utility in returning to an older assessment technique to benefit today's physics and astronomy students.

\section*{Acknowledgments}
I would like to acknowledge the reviewers, whose comments helped to improve the clarity and completeness of this paper; Dr. Jamie White, who provided the impetus for me to begin writing up and publishing pedagogical methods; the students in my classes at Colby College and Juniata College, who graciously and enthusiastically provided me with an environment to test my ideas and provided honest feedback; Swarthmore College, which provided the funding for publication; and Dr. Ruth Murray-Clay, who administered an oral exam to me for the first time that served as the nucleus for this method.

The author has no conflicts to disclose. BDZ performed all work for this paper.

\bibliography{ref}

\appendix
\section{Example questions} \label{examples}
Below are some example categories and questions for three different physics and astronomy courses. The modern physics course was taught to a class of 13 physics majors and minors (mostly sophomores), and the astronomy courses were taught to classes of 7 and 15 physics majors and minors respectively (all years). All classes were taught in physics departments at two separate small liberal arts colleges.

\subsection{Modern physics}
\begin{figure}
    \centering
    \includegraphics[width=\linewidth]{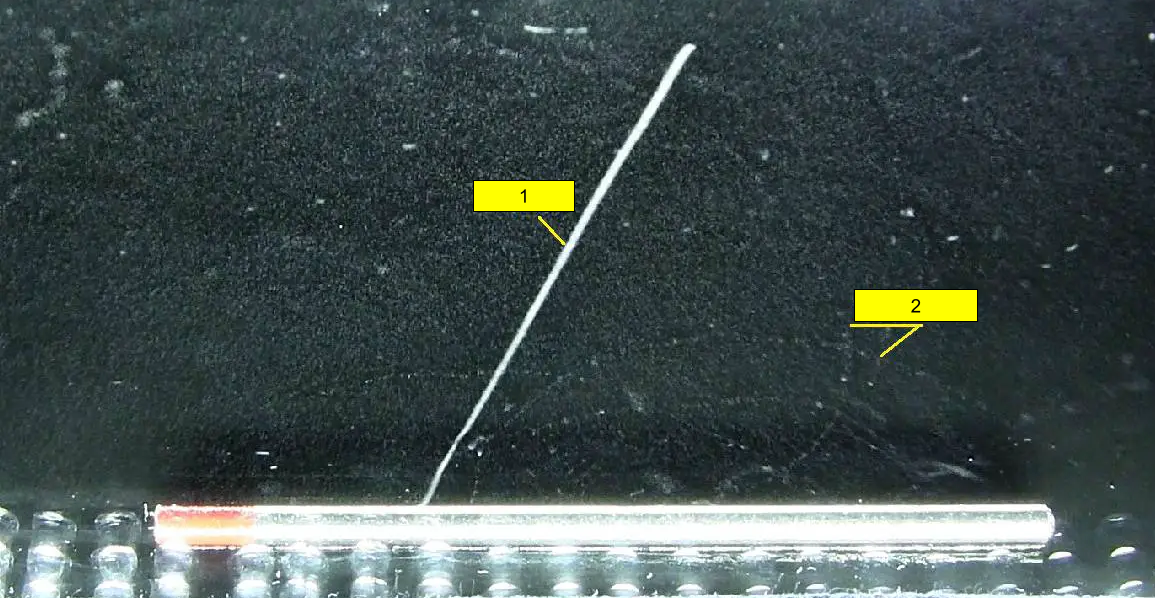}
    \caption{The cloud chamber picture for question 3 in the Cathode Ray Tube category, with a radiation source in a cloud chamber. The strong, straight trail is formed by the $\alpha$ particle labeled as $1$ and the fainter curved lines of the $\beta^-$ particles labeled as $2$. If the electron curves are too faint to be seen, they can be manually traced for the student.}
    \label{cloudchamber}
\end{figure}
\textbf{Cathode Ray Tubes:}
\begin{enumerate}
    \item Describe the construction of J.J. Thomson’s cathode ray tube experiment and why it was built the way it was.
    \item Why was the Thomson experiment unable to find the actual charge of the electron? Why was another experiment needed?
    \item A cloud chamber is a device that shows the paths of charged particles by creating the conditions for them to condense clouds as they travel through the chamber, with a magnetic field to curve the particles. The picture of the cloud chamber [Fig. \ref{cloudchamber}, provided separately to student\footnote{Adapted from Wikimedia, \url{https://commons.wikimedia.org/wiki/File:Alpha_particle_and_electrons_from_a_thorium_rod_in_a_cloud_chamber.jpg}}] has a radioactive source giving off both $\alpha$ particles and $\beta^-$ particles in a magnetic field coming out of the page. Which one is which? How do you know? Why does one curve more than the other?
\end{enumerate}

\textbf{Stefan-Boltzmann Law}
\begin{enumerate}
    \item What is the Stefan-Boltzmann law, and how would you derive it from a black-body spectrum? (You don’t have to actually do it, just say how you would do it.)
    \item Imagine that the Sun is a cube with side length $x$ rather than a sphere. What would be the formula for the total amount of power given off by the cubic Sun?
    \item In about 5 billion years, the Sun will become a red giant, and its radius will grow by a factor of about 200 while its surface temperature falls to about half its current temperature. Will it be easier or harder to look at the surface of the Sun? Will it be brighter or dimmer overall?
\end{enumerate}

\textbf{Selection Rules}
\begin{enumerate}
    \item Explain the selection rules that determine which transitions are allowed between atomic energy levels. Why do these rules exist?
    \item Draw a diagram showing all of the possible transitions that an electron in the $n=4,\ \ell=2$ state in a hydrogen atom can make, either to higher or lower energy levels.
    \item If you have ever seen the aurora borealis, you may have noticed that they often appear green on the bottom and red on the top. These two colors correspond to two different forbidden emission lines. Why do these forbidden transitions only occur high in the upper atmosphere? Which transition would you expect to be more improbable, the red line or the green line?
\end{enumerate}

\subsection{Observational and Stellar Astrophysics}
\textbf{Celestial Sphere}
\begin{enumerate}
    \item Draw a celestial sphere and label the celestial poles, the celestial equator, and the ecliptic. Explain how the RA/Dec coordinate system is defined on this sphere.
    \item The Bass Harbor Head Light lighthouse in Acadia National Park is on a stretch of coastline that extends perfectly east/west, so its view of the ocean is directly south. During what times of the year can you watch the sunrise over the ocean from the lighthouse, and during what times of the year is it blocked by the peninsula it is on?
    \item Uranus has a rotational axis that is only 7 degrees off of being aligned with the plane of its orbit. How would this change its celestial sphere compared to Earth's?
\end{enumerate}

\textbf{Magnitudes}
\begin{enumerate}
    \item Why is it easier to talk about differences in magnitudes between stars than absolute magnitudes? How do we define absolute magnitudes to get around this?
    \item How would the formula for absolute magnitude change if we defined it as the brightness at 100 pc instead of 10 pc?
    \item Say that Hipparchus’s star magnitude system went from 1 to 4 instead of 1 to 6, so our modern system of magnitudes was defined such that a 3 magnitude difference corresponded to a factor of 100 in apparent brightness instead of a 5 magnitude difference. How would this change the equation for the difference in magnitudes as a function of the ratios of fluxes?
\end{enumerate}

\textbf{Stellar Evolution}
\begin{enumerate}
    \item Draw the path that the Sun will trace on the HR diagram over the course of its lifetime, labeling each of the stages.
    \item Draw a graph of the radius of the Sun over the course of its lifetime, labeling each of the stages and explaining what causes each of the features on the graph.
    \item Which would you expect to have a higher temperature at the core, a 100 solar mass main sequence star or a 1 solar mass star on the horizontal giant branch? Why?
\end{enumerate}

\subsection{Galaxies and Cosmology}
\begin{figure}
    \centering
    \includegraphics[width=.8\linewidth]{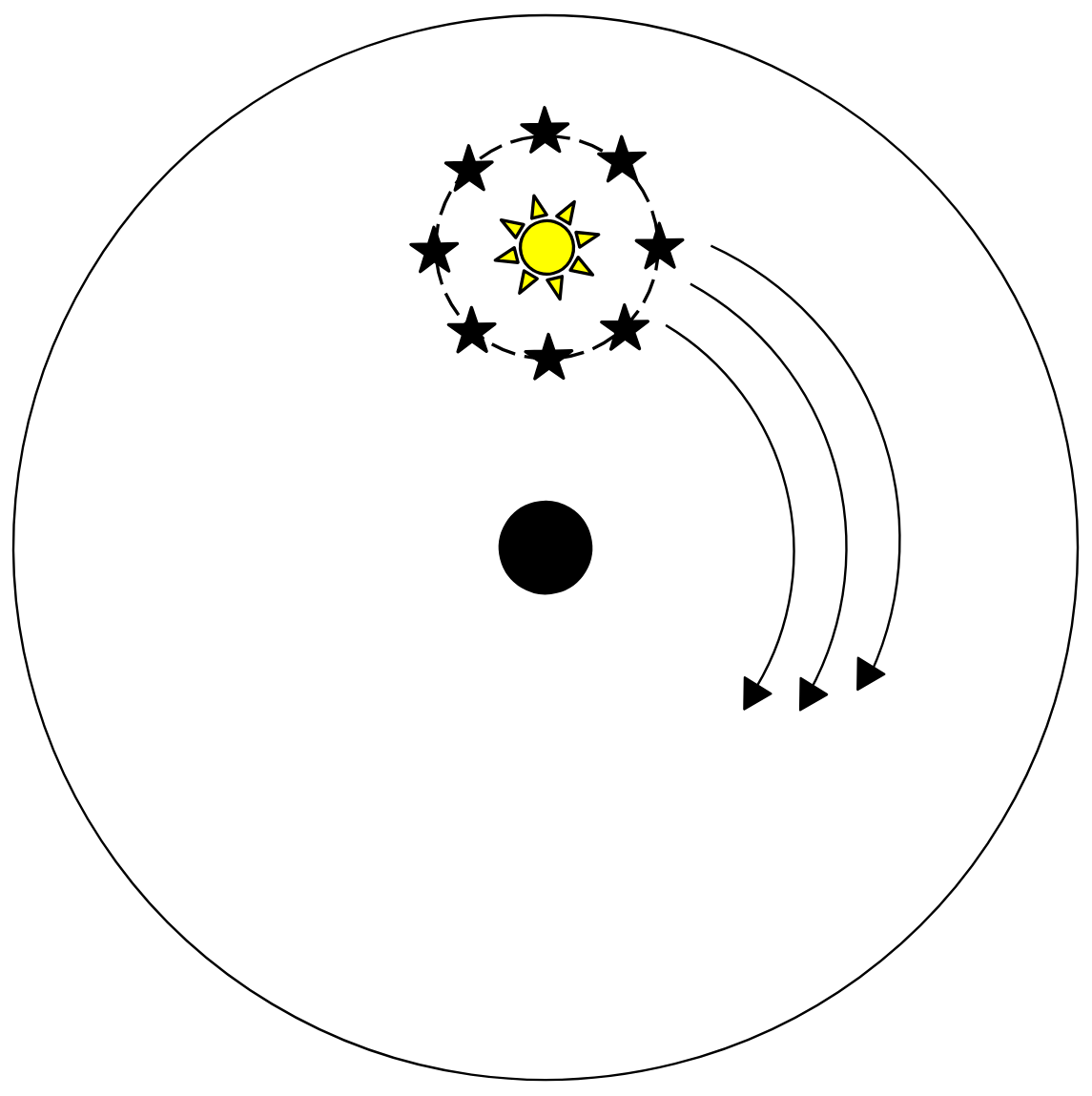}
    \caption{A schematic of a disk galaxy showing a small circle of stars around the Sun as they all orbit the center of the galaxy (used as a reference for question 2 in the Galactic Orbits category.}
    \label{orbits}
\end{figure}
\textbf{Galactic Orbits}
\begin{enumerate}
    \item Derive the formula for circular orbital velocity.
    \item Imagine there is a set of stars that are arranged in a small circle around the Sun right now so that some are slightly in front of the Sun, some are behind the Sun, some are at smaller galactic radii than the Sun, and some are at larger radii than the Sun [see Figure \ref{orbits}, which can be drawn or shown to the student]. How would you expect these stars to move relative to the Sun over time?
    \item How would you expect the result from above to change \textbf{if the Sun were at different galactic radii}? Is there any radius where you think it would be different?
\end{enumerate}

\textbf{Velocity Dispersion}
\begin{enumerate}
    \item Describe the lifetime of a star that is currently in the thick disk and how it got there.
    \item You observe a galaxy and find that its stars have a high average velocity dispersion. What does that mean for the star formation history of the galaxy?
    \item How would you expect the scale height of the thick disk to change with galactic radius?

\end{enumerate}

\textbf{Equation of State Parameter}
\begin{enumerate}
    \item How does the density of matter in the Universe relate to scale factor? Why does this make sense?
    \item Imagine we discover a new component of the Universe with $w = -2$. What would its properties be?
    \item It’s possible that we still haven’t discovered all of the components of the Universe. If there is an undiscovered component of the Universe still, what ranges would you expect its equation of state parameter to be in?

\end{enumerate}

\end{document}